\documentclass[useAMS,usenatbib]{mn2e}

\usepackage{graphicx}
\usepackage{amsmath}
\def\etal{et  al.\ }
\def\araa{{Ann.\ Rev.\ Astron.\ Ap.}}

\def\apj{ApJ}
\def\apjl{{ApJ\ (Lett.)}}
\def\apjs{{ApJ\ Suppl.}}

\def\aap{{A\&A}}
\def\mnras{{MNRAS}}

\def\prd{{Phys. Rev. D}}
\def\physrep{{Phys.~Rep.}}   

\def \xmm {\hbox{\it XMM-Newton }}

\def \etal {et al.\ }

\title[New constraints on MOND]{New constraints on MOND from galaxy clusters }
\author[E. Pointecouteau and J. Silk]{Etienne Pointecouteau$^{1}$\thanks{E-mail: etienne@astro.ox.ac.uk (EP)} and Joseph Silk$^{1}$\thanks{E-mail: silk@astro.ox.ac.hk (JS)} \\
$^{1}$Astrophysics, University of Oxford, Keble Road, Oxford OX1 3RH}

\begin{document}

\date{;}

\pagerange{\pageref{firstpage}--\pageref{lastpage}} \pubyear{2002}

\maketitle

\label{firstpage}

\begin{abstract}
  We revisit the application of Modified Newtonian Dynamics (MOND) to
  galaxy clusters.  We confront the high quality X-ray data for eight
  clusters of galaxies observed by the \xmm satellite with the
  predictions of MOND.  We obtain a ratio of the MOND dynamical mass
  to the baryonic mass of $M_m/M_b=4.94\pm 0.50$ in the outer parts
  (i.e $r\sim 0.5$~R$_{vir}$), in the concordance cosmological model
  where the predicted asymptotic ratio, if any baryons are present, is
  $7.7^{+1.4}_{-1.1}$ (at $r\geq 0.3$~R$_{vir}$).  We confirm that the
  MOND paradigm lowers the discrepancy between the binding mass and
  the baryonic mass in clusters by a factor of $\sim 1.6$ at about
  half the virial radius. However, at this radius about 80\% of
  the mass is still missing, and as pointed out by \citet{sanders03}, this
  necessitates a component of dark baryons or neutrinos in the cluster
  core. Concerning the neutrino hypothesis, application of the
  new data requires a minimum neutrino mass of $m_\nu>1.74\pm 0.34$~eV
  to fill this gap. The corresponding 2$\sigma$ lower limit of
  $m_\nu>1.06$~eV is marginally inconsistent with the current
  constraints from the cluster number counts, and from the CMB and
  large scale structure data. MOND must invoke neutrinos to
  represent the main component that account for the missing mass problem in
  clusters.
\end{abstract}
\begin{keywords}
Cosmology: observations -- Galaxies: clusters: general
\end{keywords}

\section{Introduction}

The missing mass problem in clusters of galaxies arises from the
comparison of the observed baryonic mass with the observed dynamic
mass. The baryonic mass is mainly due to the hot intracluster gas that
is well observed in X-rays via its free-free emission. The current
status of the observed gas fraction in clusters gives a fairly well
constrained value of about 12\% \citep[see][ for
instance]{grego01,allen03}.  Taking into account the stellar mass,
this makes the discrepancy between the observed dynamic mass and the
observed baryonic mass larger than a factor of $7$.

The dark matter (DM hereafter) hypothesis appears to provide a
seductive explanation of this problem. A new component of non-baryonic
matter, insensitive to all interactions but gravitation, is introduced
to fill the gap between the baryonic matter and the binding mass.
While cosmological evidence is accumulating in favour of this scenario
\citep[see for instance][]{freedman01, spergel03,tegmark04}, it is
disconcerting that the nature of the non-baryonic dark matter is
completely unknown. Of course there are many candidates of varying
degrees of detectability and plausibility \citep[eg, review by][]{bertone04}

As an alternative to dark matter scenarios, \citet{milgrom83} proposed
a modification of the Newtonian dynamics effective at galactic and
extra-galactic scales.  This modified Newtonian dynamics (MOND
hereafter) has been notably successful in explaining the discrepancy
between rotation and luminosity curves in spiral galaxies
\citep{milgrom83b,milgrom83c}, and claims other phenomenological
successes \citep[see][ for a review]{sanders02}. Given that there is
now a relativistic, Lorentz invariant generalisation of MOND that can
incorporate both gravitational lensing and cosmology \citep{bekenstein04},
it is timely to reexamine one of the few admitted failures of MOND.
The discrepancy between the baryonic mass and the dynamical mass in
clusters of galaxies is perhaps foremost among the issues that MOND
has yet to convincingly address.

The first confrontation of X-ray observations of clusters with MOND
\citep{gerbal92} emphasised, despite some minor controversy
\citep{milgrom93,gerbal93}, the difficulties faced by MOND in passing
the cluster test. The problem was revisited by \citet{sanders94,
  sanders99} and ended in a remaining discrepancy of a factor of 2-3
between the baryonic and the MOND masses.  More recently,
\citet{aguirre01} discussed observational evidence for three clusters
for which the observed discrepancy is about 1-5 within 1~Mpc and is
boosted to a factor $\sim 10$ within the central 200~kpc, further
weakening the reliability of the MOND paradigm.  However
\citet{sanders03} responded with an update of his earlier work,
introducing an added \emph{ad hoc} dark component at the cluster
centre to reduce the discrepancy to only a factor of 1-3 overall in
the cluster.

Some other tests have also been carried out using gravitational
lensing data.  They have also pointed out the difficulties faced by
MOND at the cluster scale \citep{gavazzi02,clowe04}.

In this paper, we test the MOND prediction in terms of dynamical mass
with respect to the observed baryonic mass, by basing our work on high
quality X-ray data recently obtained from observations with the \xmm
satellite for ten nearby relaxed clusters
\citep{pointecouteau05,arnaud05}.  Section~2 presents the context and
the formalism within which the mass of clusters is derived in the MOND
case.  In section~3 we present the data and the working
framework we adopt to compute the observed MOND masses. The
comparison of the MOND mass with the Newtonian dynamic mass and the
baryonic mass is developed in Sec.~4.  The results are discussed with
respect to previous studies in Sec.~5. Our new result is that, in
addition to the known problem in the cluster cores, there is a
significant discrepancy in the outer cluster, where an additional
\emph{ad hoc} dark component is needed to rescue MOND. We
investigate the neutrino hypothesis for such a possibility.

Unless mentioned otherwise, we choose to work in a concordance
model, using as cosmological parameters: $\Omega=1$, $\Omega_m=0.3$,
$\Omega_\Lambda=0.7$ and $H_0=70$~km/Mpc/s (referred as LCDM).

\section{MOND dynamical masses \label{sec:mm}}
 
For a spherical system in hydrostatic equilibrium, the density and the
temperature distributions are connected via the equation of hydrostatic
equilibrium, so that the dynamical mass of a spherical system can be
expressed as :
\begin{equation}
  M_d(r) = - \frac{kT\ r}{{\rm G} \mu m_p}   \left[ \frac{d \ln{n_{\rm
          g}}}{d \ln{r}} + \frac{d \ln{T}}{d \ln{r}} \right]
\label{eq:he}
\end{equation}
where G is the gravitational constant, $m_p$ is the proton mass and
$\mu=0.609$. $M_d(r)$ is the dynamical mass enclosed within a sphere
of radius $r$. Here, $n_g(r)$ and $T(r)$ are the density and temperature
radial distributions.

In the framework of modified Newtonian dynamics, the gravitational
acceleration $a$ is linked to the Newtonian acceleration $g$ as follows:
\begin{equation}
a\, f(a/a_0)=g
\label{eq:amond}
\end{equation}
where $a_0$ is assumed to be a universal constant of acceleration in
MOND. Its value as determined from the rotation curves of galaxies is
about $10^{-8}$~cm/s$^{2}$.  In the following, we adopt the value of
$0.8\times 10^{-8}$ used by \citet{sanders99} (hereafter S99 --
referring to \citealt{begeman91}). The transition between the Newtonian
and the MOND regimes is expressed through the functional $f(x)$, $x$
being $a/a_0$, that is also derived from the application of MOND to
the rotation and luminosity curve of galaxies:
\begin{equation}
f(x)=\frac{1}{\sqrt{1+x^{-2}}}
\label{eq:fx}
\end{equation}

We now quote Eq.~9 from S99, that gives the MOND gravitational
acceleration $a$, quantity that decreases with increasing radius:
\begin{equation}
  a(r) = - \frac{kT}{\mu m_p\ r}   \left[ \frac{d \ln{n_{\rm
          g}}}{d \ln{r}} + \frac{d \ln{T}}{d \ln{r}} \right]
\label{eq:acc}
\end{equation}

Eq.~8 from S99  relates the dynamical MOND mass, $M_m$, to
the dynamical Newtonian mass, $M_d$:
\begin{equation}
M_m(r) = \frac{M_d(r)}{\sqrt{1+(a_0/a(r))^2}}
\label{eq:mond}
\end{equation}
Further details of the MOND formalism can be found, for instance, in S99,
\citet{sanders02} and \citet{sanders03}.

According to Eq.~\ref{eq:he} and Eq.~\ref{eq:acc}, we can express the
ratio of the Newtonian dynamical mass to the MOND mass as:
\begin{equation}
\frac{M_d}{M_m}(r) = \sqrt{1+\left(\frac{a_0}{{\rm G}}\frac{r^2}{M_d}\right)^2}
\label{eq:ratio}
\end{equation}
In the following we refer to the Newtonian dynamical mass, $M_d(r)$,
as the dynamical mass, and to the dynamical mass in modified Newtonian
dynamics, $M_m$, as the MOND mass.

\section{Working framework \label{sec:work}}
\subsection{The X-ray data}

Recently, \citet[ hereafter PAP05]{pointecouteau05} have published the
observed mass profiles for a set of ten nearby ($z\leq 0.15$) and
relaxed galaxy clusters observed with the \xmm satellite.  These
clusters, chosen for their regular X-ray morphologies, have dynamical masses 
determined  through
the hypothesis of spherical symmetry and the use of Eq.~\ref{eq:he}.
They cover a large temperature range of $[2-9]$~keV, corresponding to
a dynamical mass range of $[1.2-12.0]\times 10^{14}\rm M_\odot$ enclosed within
a radius $R_{200}$ (i.e a radius encompassing 200 times the
critical density of the Universe at the cluster redshift), and a range
of radii between 0.01 to 0.7 in units of $R/R_{200}$.

From the initial sample of ten clusters, we kept only eight for the
present work: A1991, A2717, A2597, A1068, A478, A1413, PKS0745, A2204.
We excluded A1983 and MKW9. First they are the only two clusters of
the sample that are not observed up to 0.5~$R_{200}$. Indeed, we need
to keep to the observed radial range of each cluster to avoid any
extrapolation, and to derive reliable observational constraints.
Meanwhile MWK9 has also the most disturbed morphology of the 10 clusters
\citep{pratt05}.  Both also exhibit unexpectedly low gas fractions that
may turn them into outliers in terms of the average gas fraction in
clusters \citep{grego01,allen03}.  As our sample is quite small,
excluding those two clusters may avoid any poissonian bias on the
computation of the intrinsic dispersion for the ratio of the MOND mass
to the baryonic mass.  However, at the scale of our sample, this also
may slightly bias the results in favour of MOND predictions, for which
low gas fraction systems will be difficult to explain (as it increases
the discrepancy between the observed baryonic mass and the dynamical
mass).

We used the observed density and temperature profiles derived by PAP05
to compute the observed ratios of the dynamical mass to the MOND mass
and of the MOND mass to the baryonic mass.

\subsection{Scaled MOND profiles \label{sec:sc}}
In hierarchical structure formation, the virialized part of clusters
at a given redshift is encompassed within a sphere with a radius
corresponding to a fixed density contrast ($\sim 200$) with respect to
the critical density of the Universe. $R_{200}$ is thus considered to be
the virial radius of the cluster \citep[see][ and references
therein]{arnaud04}.

Thus to cross-compare the internal structure of clusters, one has to
look at quantities scaled according to $R_{200}$.  As an extension, to
compare the MOND and the dynamical masses as a function of radius, we
choose to scale both quantities according to the virial radius (i.e
$R_{200}$). For each cluster, we used the value of $R_{200}$, reported
by PAP05 for a LCDM cosmology (see their table~2), obtained from the
fit of NFW profiles \citep{navarro97} over the observed mass profiles.
In the framework of MOND structure formation, \citet{knebe04} have
shown that the most massive galactic halos formed in their numerical
simulations could be fitted by a NFW profile.  We could thus fit each
MOND mass profile with a NFW profile to derive the corresponding
characteristic scaling radius $R_{200}^M$. However, for a given
cluster, the values of $R_{200}$ and $R_{200}^M$ are likely to be
different.  Scaling $M_d(r)$ and $M_m(r)$ respectively with $R_{200}$
and $R_{200}^M$ will lead us to compare these two profiles at
different physical scales.  For our purpose, working with scaled radii
is required by the use of a sample of clusters with different masses,
but is not required to compare the MOND and dynamical masses for a
given cluster.  A single scaling has to be adopted for each cluster.We
thus adopt $R_{200}$ as a scaling radius for both dynamical and MOND
profiles.

 \begin{figure}
\includegraphics[width=8.5cm]{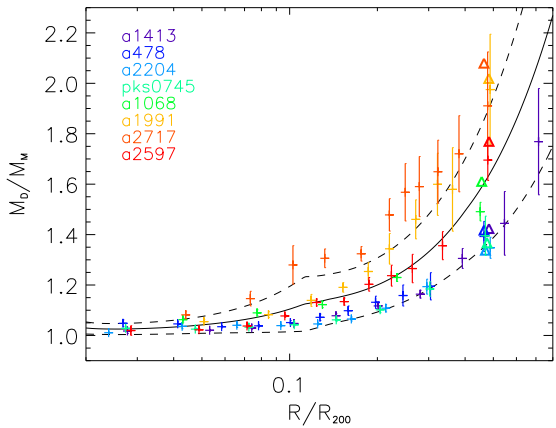}
\caption{Observed scaled profiles for the ratio of the dynamical mass
  to the MOND mass. The average profile over the sample is
  overplotted (solid black line) together with its 1$\sigma$
  dispersion (dashed lines). The triangles mark the estimate from
  the self-similar hypothesis for each cluster (see
  Sect.~\ref{sec:sc}). }
\label{fig:f1}
\end{figure}

\subsection{The baryonic  mass  in clusters\label{sec:star}}

The observational constraints on the gas fraction in clusters strongly
favour an asymptotic value of about 10-12\% beyond $0.3$~R$_{200}$
\citep{grego01,allen03}, in a LCDM cosmology.  For the sample used
here, the average gas fraction at a density contrast of $\delta=1000$
(i.e at $r=0.47\pm 0.02$~R$_{200}$ -- see APP05) is $<f_{gas}>=0.11\pm
0.02$.
We adopted this value in the following, as the average asymptotic
value for the gas fraction.

We now can further investigate the issue of the stellar mass.  For the
stellar mass, S99 used a correlation derived from the early work of
\citet{david90} between the luminosity and the gas masses in clusters
leading to: $f_{\star}=M_\star/M_{gas}\simeq 0.7 (kT/{\rm keV})^{-1}
h^{3/2}$. We used this estimate when working in CDM cosmology, making
use of the spectroscopic temperature measured between
$[0.1,0.5]$~R$_{200}$ (see APP05).  Nevertheless, more recent work
allows us to revise this crude estimate of $f_\star$.  For example,
\citet{lin03} derived a variation of $f_\star$ with the total mass,
and thus with the temperature. From their equation $10$ and the values
of $M_{500}$ derived by APP05 (see their table~1), we obtain an
average value of $<f_\star>=0.12\pm0.03$ (values ranging from 0.1 to
0.17). In related work, \citet{voevodkin04} obtained consistent, but
slightly higher, values, and concluded that the stellar mass
represents about 15\% of the gas mass.  We adopted this value of
$f_\star=15$\% for our sample.  The baryon fraction will then be
$f_b=0.13\pm 0.02$.  In order to cope with the baryonic mass in
clusters, the MOND mass has then to be $7.7_{-1.0}^{+1.4} $ times
lower than the dynamical mass once the baryonic mass reaches its
asymptotic value (i.e $r> 0.3$~R$_{200}$).

\section{Matching the baryon contents of clusters \label{sec:bar}}

In an initial approach, taking into account the self-similar nature of the
cluster population, we can express the ratio $M_d/M_m$ as a function
of $kT$. We make use of the scaling relations between the mass and the
temperature: $M=A(z)\ T_\delta^\alpha$, and between the mass and the
radius $M=4\pi/3\ \delta\, \rho(z)\, R_\delta^3$. Then Eq.~\ref{eq:ratio}
can be rewritten as follows:
\begin{equation}
\frac{M_d}{M_m}(r) = \sqrt{1+B(z) T_\delta^{-2\alpha}}
\label{eq:ratsim}
\end{equation}
where $B(z)=(3a_0/4\pi \delta\, \rho_c(z))\, A(z)^{-1}$, and $T_\delta$
is the temperature at  density contrast $\delta$.

Using the $R-T$ relation published by APP05 at $\delta=1000$ (a
density contrast corresponding to observed radii for all eight
clusters) we estimated from the above equation $M_d/M_m=1.63\pm 0.29$,
which leads to $M_m/M_b\simeq 4.7$.

We can reverse the computation and estimate the value of $a_0$ needed
to reach an equality between the MOND mass $M_m$ and baryonic mass
$M_b$. We obtain $a_0=(4.75\pm 1.24)\times 10^{-8}$~cm/s$^2$ at
$\delta=1000$ using the eight clusters.
~\\

From the mass profiles derived by PAP05 in a LCDM cosmology, we
computed the observed ratio $M_d/M_m$ from Eq.~\ref{eq:ratio} for the
eight clusters of the sample. The profiles (scaled to $R_{200}$) are
shown in Fig.~\ref{fig:f1}. The estimates from the self-similar
hypothesis are shown for $\delta=1000$ as the observed triangles, and
match the computed ratio very well. 

With the previously given notations and Eq.~\ref{eq:ratio}, we can
express the ratio of the MOND mass to the baryonic mass as:
\begin{equation}
\frac{M_m}{M_b}(r) = \left[f_{gas}(1+f_\star)\sqrt{1+\left(\frac{a_0}{{\rm G}}\frac{r^2}{M_d}\right)^2}\right]^{-1}
\label{eq:mmvsmb}
\end{equation}
In Table~\ref{tab:ratio}, we report the average baryon fraction values
and the average $M_m/M_b$ and $M_d/M_m$ ratios for the measured clusters
at a given radius.
The computation were done at the following radii: the physical radius of
750~kpc in order to directly compare with S99, and the radii
corresponding to the density contrasts of $\delta =1000$ and
$\delta=15000$ (i.e $0.47\pm 0.02$~$R_{200}$ and $0.10\pm
0.01$~R$_{200}$ average over the sample). Those two radii mark the
boundaries of the radial range over which the observational
constraints are especially well tied down (see PAP05, APP05). 

From Fig.~\ref{fig:f1}, it is obvious that the ratio $M_d/M_m$
decreases, thus the discrepancy between the MOND and the baryonic mass
increases, towards the centre of clusters as the acceleration returns
to the Newtonian regime. At $\sim 0.1$~R$_{200}$, the ratio is
$1.08\pm 0.06$.  The discrepancy also increases for massive (i.e hot)
systems. Indeed if we only consider the 5 hot clusters\footnote{i.e
  A1413, A478, A2204, PKS0745, A1068} 
we obtain at $\delta=1000$ a ratio of $M_d/M_m=1.43\pm 0.08$, an even
smaller value. The corresponding value of $a_0$ to match the baryonic
mass has to be $a_0=(5.58\pm 0.63)\times 10^{-8}$~cm/s$^2$.

\section{Discussion \label{sec:dis}}

\subsection{MOND as a stand alone solution \label{sec:alo}}
For a direct comparison with S99, it is interesting to performed the
computation at the radius of 750~kpc with $\Omega_m=1$ and
$H_0=50$~km/s/Mpc. For the five clusters observed beyond 750~kpc
(i.e. the 5 hot clusters), the average gas fraction is then $0.19\pm
0.02$ and the derived ratio of the MOND mass to the baryonic mass is
$M_m/M_b=3.84\pm 0.33$.  For the same cosmological setup, S99 claimed
that the MOND context reduces the discrepancy with the baryonic mass
down to an ``acceptable'' factor of $\sim 2$.  However, our value of
$3.8$ is significantly higher, and does not even agree with a factor
of 2 within a 3$\sigma$ limit. 

As we are working with a sample of nearby clusters, the most important
cosmological parameter is the Hubble constant.  Indeed, the baryon
fraction in clusters scales with $h^{-3/2}$, thus decreases with
increasing values of $H_0$.  Current observational evidence strongly
favours a flat Universe with a low matter density $\Omega_m\sim 0.3$,
a component due to dark energy which can be represented by a positive
cosmological constant $\Lambda$, and a high value for the Hubble
constant ($H_0\sim 70$~km/s/Mpc) \citep[see][ for
instance]{freedman01, spergel03,tegmark04,seljak05}. In such a
universe (i.e.  LCDM) $M_d/M_m$ decreases to $1.63\pm 0.31$ ($1.08\pm
0.06$) at $\sim 0.5$~R$_{200}$ ($\sim 0.1$~R$_{200}$), so that the
discrepancy is increased between the MOND dynamic mass and the
baryonic mass.  The corresponding ratio of the MOND mass to the
baryonic mass is now $4.94\pm 0.50$ ($10.6\pm3.77$).  This is more
than a factor of two above the value derived by S99.  The evidence is
confirmed if we only consider the hot systems. Indeed, for clusters
with ($kT>3.5$~keV), we derive $M_d/M_m=1.43\pm 0.08$ at $\sim
0.5$~R$_{200}$, which makes the ratio of the MOND mass to the baryonic
mass $\sim 5.10\pm 0.56$. Thus in all cases, within a 3$\sigma$ (i.e.
99\% confidence) the ratio $M_m/M_b$ in a MOND cosmology will be
greater than 3.4, making MOND unable to fully overcome the missing
mass problem in clusters.

The case of A1413 is even more eloquent. Indeed, this cluster has been
observed beyond $R_{500}$ (i.e. a physical radius of 1129 kpc
corresponding to $\sim 0.7~$R$_{200}$ -- \citealt{pratt02}), a radius
that conservatively is often taken as the outer bound of the
virialized part in clusters. Moreover, for this cluster the \xmm
results agrees very well with the results derived from the {\it
  Chandra} observations, especially in term of the shape of the
temperature profile \citep{vikhlinin05}. At the radius of $R_{500}$,
the measured gas fraction is $f_{gas}=0.15\pm 0.01$, which makes the
baryon fraction $0.17\pm 0.01$ (see Sect.~3.3).
We derived a corresponding ratio $M_m/M_b=3.6\pm 0.7$.  Such an
observed discrepancy measured within a radius closing the virial
radius puts a very tight constraint on the MOND paradigm.

In fact, S99 relies on hypotheses that together might have biased its
results down towards an optimistic value: (i) the use of a
$\beta$-model to describe the gas distribution down to 750~kpc
contributes to overestimate the gas mass at large radii. Moreover
the spatial resolution of the current X-ray data has ruled out the
$\beta$-model as a fitted representation of the observed X-ray surface
brightness profiles (especially in nearby clusters -- see PAP05 for
instance); (ii) The use of physical radius as a working radius could
have induced biases as clusters of very different masses are then
compared at various different density contrasts; (iii) The
isothermality of the intra-cluster medium was forced by the single
overall temperature measurements available for the considered sample.
However, in contrast we have made use of accurately measured temperature
profiles.

If we consider the problem in terms of the value of the characteristic
MOND acceleration, we derive values of $a_0$ that are $5$ times larger
(and still $3.5$ times larger within the 3$\sigma$ limit) than the
value derived from the rotation curves of galaxies, and are thus
unacceptable values.

\begin{table}
  \caption[]{Ratios of the dynamical mass to the MOND mass and of the MOND mass to the baryonic mass.\label{tab:ratio}}
\begin{flushleft}
\begin{center}
\begin{tabular}{llccc}
  \hline
  \hline
  Radius & $f_b$ & $M_d/M_m$ & $M_m/M_b$ & $N_c\, ^{a}$ \\
  \hline
  $R=750$~kpc  & $0.13\pm 0.01$ & $1.34\pm 0.14$ & $5.61\pm 0.55$ & 5 \\
  $\delta=15000$ & $0.09\pm 0.03$ & $1.08\pm 0.06$ & $10.6\pm 3.77$ & 8 \\
  $\delta=1000$  & $0.13\pm 0.02$ & $1.63\pm 0.31$ & $4.94\pm 0.50$ & 8 \\
  $\delta=1000\, ^{b}$ & $0.14\pm 0.02$ & $1.43\pm 0.08$ & $5.10\pm 0.56$ & 5 \\
  \hline
\end{tabular}
\end{center}
(a) Number of clusters used.
(b) Ratio for the hot clusters only (i.e $kT >3.5$).
\end{flushleft}
\end{table}

\subsection{Adding a dark component \label{sec:dc}}
A last alternative to rescue MOND is to invoke a non-luminous
component at the centre of clusters, as suggested recently by
\citet{sanders03} (S03 hereafter). This author introduced this dark
component on an \emph{ad hoc} basis to explain the huge observed
discrepancy within the cluster centre \citep{aguirre01}. Indeed, the
baryon fraction is lower in the inner parts, so that the ratio to the
MOND mass increases. With respect to our sample, if such a component
was to exist it would have to account for $81$\% ($\pm 4$\%) of the
total mass at $\sim 0.5$~R$_{200}$.  In other terms, this means that
MOND just reduces the missing mass problem in clusters by about 20\%
(at half the virial radius) compare to a standard DM scenario, but
does not solve it. Therefore, some exotic dark component has to be
added to fill the remaining gap of 80\%. This component will become
the main explanation for the missing mass, and will thus draw a
cosmological setup that will be closer to a mixed DM scenario than to
a dominating MONDian scenario.

It is highly unlikely that 80\% of the missing mass in cluster can be
due to hidden baryons. Indeed, if atomic or molecular hydrogen exist
within the intra-cluster medium, it will have to face its thermal
conditions. Such a component could only account for a small
fraction of the thermal baryons. The relativistic populations are also
known to be a minor component in terms of mass, as otherwise they will
be expected to produce a strong hard X-ray signal, and stronger radio
emissions than those currently observed.

Nevertheless, we can follow the suggestion by S03 to consider, as
potential candidates for this dark component, massive neutrinos that
aggregate at cluster scale.  In the MOND case, according to S03, the
needed neutrino mass per neutrino flavour needed for MOND to cope with
the observed baryonic mass  is about 2~eV.
Further assuming a constant density sphere for his dark component and
taking into account the phase space density limit for neutrinos, S03
derived an upper limit for the neutrino density after their collapse
and accretion within structure of: $\rho_\nu\leq (4.8\times 10^{-24})
(m_{\nu}/2\textrm{eV})^4 (T_{\textrm{keV}})^{3/2}$ kg m$^{-3}$.  We
make use of this limit on the neutrino density, and we use the
spectroscopic temperatures measured for each cluster of our sample
between 0.1 and 0.5~$R_{200}$ (see PAP05).  From our eight clusters,
it is possible to compute the needed neutrino mass to equate the
missing mass at a given radius (i.e. $M_m(r)-M_b(r)$) with the
contribution of the massive neutrino to the cluster total mass.  S03
hypothesis of the neutrino accretion mainly concerned the central
parts of clusters. In our study, to explain the $\sim 80$\% of missing
mass in MOND, we add to extend the radius of the neutrino sphere down
to $R_{1000}$. The minimum neutrino mass then required is $m_\nu>
1.74\pm 0.34$~eV . This is a strongly constraining value for the
neutrino mass, which makes the lower bound for the neutrino mass
becomes $\sim 1.06$~eV, within a 2$\sigma$ limit (i.e. 95\%
confidence).

To date particle physics experiments (single and double beta-decay,
neutrino oscillations measurements) lead to a wide range of upper
limit for the neutrino mass going from $\sim0.8$~eV to $\sim3$~eV
\citep{fogli04}. Those constraints directly apply to MOND, and they
are consistent with our lower limit taking into account their large
variation.
If to date the constraints derived from astrophysical data seems to be
tighter, they
may be considered as irrelevant in a MOND framework (i.e. $\Omega_\nu
h^2=\sum{m_{\nu_i}}/94\textrm{eV}$) as they are derived within the
framework of a standard cosmological DM model for structure
formations.  Nevertheless, within the cosmological MOND+neutrino setup
we end up with, MOND plays a minor role, where the neutrinos are the
major component to explain the missing mass in the Universe.  We can
thus reconsider the astrophysical constraints in this context.  As the
number of clusters is linked to the matter content of the Universe,
thus to $\Omega_m$ and $\Omega_\nu$, the cluster number counts can be
used to give an upper limit on the neutrino mass
\citep{kahniashvili05,elgaroy05}.  With the following cosmological
setup: $\Omega_m=0.3$, $h=0.7$, $n=1$ and $0.7<\sigma_8<1.1$ (to
account for the variation in $\sigma_8$ determinations -- see for
instance \citet{tegmark04,seljak05b}), current cluster number counts
lead to an 2$\sigma$ upper limit of $m_\nu<0.8$~eV
\citep{fukugita00,allen03}. A value marginally incompatible with the
lower bound was derived here.  This inconsistency is even stronger with the
2$\sigma$ upper limit of $\sim 0.3$~eV derived from the combination of
the WMAP, 2dFGRS and SDSS data
\citep{spergel03,tegmark04,seljak05,elgaroy05}.  However, one has to
keep in mind that there is quite a large dispersion between
those upper limits \citep{elgaroy05}, that may just leave enough room
for the MOND+neutrino hypothesis to stand as a reliable paradigm.

\section{Conclusion \label{sec:con}}

We have revisited the case of the galaxy cluster scale in the
framework of the MOND paradigm. We based our study on a set of
consistent and recent high quality X-ray data obtained with the last
generation of X-ray satellites (\xmm).  For eight nearby and relaxed
clusters, we confirmed that MOND alone is not able to explain the
missing mass problem in clusters of galaxies.  Almost 80\% of the mass
is still missing at half the virial radius of clusters in a MOND
Universe.  Thus MOND is not the main solution to the missing mass
problem at cluster scales.

Indeed, it undeniably requires extra exotic DM to survive. In this
context we have investigated the hypothesis of massive neutrinos
distributed as a sphere of constant density up to half the virial
radius of galaxy clusters.  Under this hypothesis, we derived a very
tight observational lower bound for the neutrino mass, $m_\nu>1.06$~eV
(95\% confidence), which is marginally inconsistent with the
constraints from the cluster number counts and from the CMB and LSS
constraints. The consistency with the current direct experiments to
measure the neutrino mass is spoiled by their associated large
uncertainties. The perspectives of dedicated CMB and Large-scale
structure experiments in coming years will bring definitive
constraints on $m_\nu$ in the astrophysical context, and should
definitively settle the case of the MOND paradigm.

Note that recently \citet{skordis05} have been studying the formation
of structure in the relativistic MOND framework (i.e.  the Bekenstein
theory -- see \citet{bekenstein04}).  To reproduce the observed
angular power spectrum of the CMB, those authors appeal to massive
neutrinos with $m_\nu\simeq 2$~eV.  So unless the main fraction of the
baryonic content of clusters remains hidden from the current
observations, in any case a large neutrino mass (or another exotic
massive candidate) is needed to promote MOND as a reliable paradigm.
As such, massive neutrinos would then become a major component of the
Universe matter content. This will turn the cosmological framework
more into a mixed DM cosmology than into a MONDian cosmology.

\section*{Acknowledgements}
We thank the anonymous referee for her/his fruitful comments. We are
grateful to Monique Arnaud, Mathieu Langer, Adi Nusser and Gabriel
Pratt for useful discussions and comments. EP also thanks Monique
Arnaud for software issues and  Adi Nusser for enlightenments on
the basis of MOND. EP acknowledges the financial support of the
Leverhulme Trust (UK).


\end{document}